\documentclass[prl,twocolumn,showpacs,amsmath,amssymb]{revtex4-1}

\usepackage{graphicx}
\usepackage{amsfonts}
\usepackage{amsmath}
\usepackage{amssymb}
\usepackage{color}
\usepackage{bm}
\usepackage{bbm}
\usepackage{enumerate}
\usepackage{color}
\graphicspath{figures} 
\usepackage{amsfonts, amsmath, amsthm, amssymb} 
\usepackage{mathtools}
\usepackage{array}
\usepackage[colorlinks,bookmarks=false,citecolor=blue,linkcolor=red,urlcolor=blue]{hyperref}

\usepackage[colorlinks,bookmarks=false,citecolor=blue,linkcolor=red,urlcolor=blue]{hyperref}
\usepackage[usenames,dvipsnames,svgnames,table]{xcolor}
\graphicspath{figures} 

\newcommand{\be}{\begin{equation}}
\newcommand{\bee}{\begin{equation*}}
\newcommand{\ee}{\end{equation}}
\newcommand{\eee}{\end{equation*}}
\newcommand{\bearre}{\begin{eqnarray*}}
\newcommand{\eearre}{\end{eqnarray*}}
\newcommand{\bearr}{\begin{eqnarray}}
\newcommand{\eearr}{\end{eqnarray}}

\begin{document}

\title{
	Quantum critical metrology
}

\author{Ir\'en\'ee Fr\'erot$^1$\footnote{Electronic address: \texttt{irenee.frerot@ens-lyon.fr}},
and Tommaso Roscilde$^{1,2}$
 }

\affiliation{$^1$ Univ Lyon, Ens de Lyon, Univ Claude Bernard, CNRS, Laboratoire de Physique, F-69342 Lyon, France}
\affiliation{$^2$ Institut Universitaire de France, 103 boulevard Saint-Michel, 75005 Paris, France}
\date{\today}


\begin{abstract}
Quantum metrology fundamentally relies upon the efficient management of quantum uncertainties. We show that, under equilibrium conditions, the management of quantum noise becomes extremely flexible around the quantum critical point of a quantum many-body system: this is due to the critical divergence of quantum fluctuations of the order parameter, which, via Heisenberg's inequalities, may lead to the critical suppression of the fluctuations in conjugate observables. Taking the quantum Ising model as the paradigmatic incarnation of quantum phase transitions, we show that it exhibits quantum critical squeezing of one spin component, providing a scaling for the precision of interferometric parameter estimation which, in dimensions $d \geq 2$, lies in between the standard quantum limit and the Heisenberg limit. Quantum critical squeezing saturates the maximum metrological gain allowed by the quantum Fisher information in $d=\infty$ (or with infinite-range interactions) at all temperatures, and it approaches closely the bound in a broad range of temperatures in $d=2$ and 3. This demonstrates the immediate metrological potential of equilibrium many-body states close to quantum criticality, which are accessible \emph{e.g.} to atomic quantum simulators via elementary adiabatic protocols.  
\end{abstract}
 \maketitle


\textit{Introduction.} Observables in extended physical systems (classical or quantum in nature) are affected by intrinsic uncertainty, which typically results from an extensive number of uncorrelated, microscopic local contributions. As a consequence the squared uncertainty scales linearly with system size, in compliance with the central limit theorem. Yet collective phenomena, such as phase transitions, may lead to the appearance of sizable correlations among the constituents, leading to the breakdown of the central limit theorem and to super-extensive scaling of fluctuations, which clearly aggravates the uncertainty of the corresponding observable. Yet in quantum systems uncertainties of non-commuting observables $A$ and $B$ may play complementary roles as they obey the Heisenberg's inequality ${\rm Var}(A) {\rm Var}(B) \geq |\langle [A,B] \rangle|^2/4$ (where ${\rm Var}(A) = \langle A^2 \rangle - \langle A \rangle^2$ and $\langle ... \rangle = {\rm Tr}(\rho ...)$ denotes the average on the state $\rho$ - pure or mixed - of the system). In the following we shall focus on the physically relevant situation in which $A$, $B$ and $[A,B]$ are macroscopic observables.  As a consequence of Heisenberg's inequality, the critical increase of fluctuations of $A$ leads to a suppression of the lower bound for the fluctuations of $B$. The reduction of a lower bound is hardly constraining for the actual behavior of fluctuations, but it may be so for quantum states realizing minimal (or close to minimal) uncertainty, namely (nearly) saturating Heisenberg's inequality. 

In this paper we show that this counterintuitive mechanism of \emph{critical suppression of fluctuations}, by which the scaling of ${\rm Var}(B)$ becomes sub-extensive when the one of {\rm Var}(A) becomes super-extensive, is indeed at play at a \emph{quantum critical point} (QCP) \cite{sachdevbook} occurring in the ground state of quantum many-body systems, implying that a QCP generically allows one to tune the quantum noise of macroscopic observables to extraordinarily low values. The redistribution of quantum noise among observables is known in the quantum-optics and atomic-physics literature as \emph{squeezing} \cite{squeezingbook,Maetal2011,Pezzeetal2016}: in the context of quantum spin systems (modeling electronic/nuclear spins in solids, or the internal states of atomic ensembles), spin squeezing \cite{Maetal2011} has both a fundamental meaning as a manifestation of entanglement \cite{Sorensenetal2001, SorensenM2001, Tothetal2009}; as well as an immediate application in the context of quantum metrology, leading to a fundamental gain in interferometric quantum parameter estimation \cite{Winelandetal1994}. In particular we show here that paradigmatic spin models of quantum phase transitions (QPTs) exhibit not only quantum-critical spin squeezing at the zero-temperature QCP \cite{Vidaletal2004,DusuelV2004} - which generically implies the sub-extensive scaling of the variance of one observable; but that squeezing is manifest in a \emph{broad region of the finite-temperature phase diagram} around the QCP, making it of interest to realistic metrological protocols. Even more importantly, for sufficiently high dimensions we show that equilibrium squeezing nearly saturates the maximum metrological gain dictated by the quantum Fisher information \cite{PezzeS2014,TothA2014}, demonstrating that a metrological protocol which exploits the equilibrium spin squeezing of thermal states in the vicinity of a QCP is (nearly) optimal. These results pave the way for a quantum-technological use of the enhanced entanglement and quantum correlations associated with quantum critical phenomena. 

 Before entering into the core of our paper, we would like to stress that our discussion of the metrological use of phase transitions is very different from that offered previously in the literature on Hamiltonian parameter estimation -- see Refs.~\cite{Zanardietal2008,Invernizzietal2008,Salvatorietal2014,Skotiniotisetal2015,Mehboudietal2015,Mehboudietal2016} for some representative examples.  The main focus of this literature is the distinguishability among equilibrium states, which becomes maximal around a phase transition (be it of thermal \cite{QuanC2009} or quantum \cite{ZanardiP2006} nature) allowing for an optimal estimation of the parameter driving the transition itself (an external magnetic field, temperature, etc.). On the other hand, our study focuses on equilibrium many-body states used as input states of interferometers (namely unitary transformations parametrized by a phase $\phi$ \cite{PezzeS2014,Pezzeetal2016}), and their augmented ability to estimate the interferometric phase in the presence of quantum correlations.    
 
\emph{Model.} Throughout this paper we focus our attention on a paradigmatic spin model of quantum phase transitions, namely the transverse-field Ising (TFI) model, whose Hamiltonian on finite-dimensional systems reads
\begin{equation}
{\cal H} = - J\sum_{\langle ij \rangle}  S^z_i S^z_j - \Gamma \sum_i S_i^x
\end{equation} 
where $S_i^{\alpha}$ are $S=1/2$ quantum spins, the sums run on nearest-neighboring pairs and sites (respectively) of a $d$-dimensional hypercubic lattice, containing $N=L^d$ sites, and $J>0$. In the special case of $d=\infty$ (or an infinite-connectivity model), the Hamiltonian takes rather the form
\begin{equation}
{\cal H} = - \frac{J}{N} \sum_{i<j}  S^z_i S^z_j - \Gamma \sum_i S_i^x~.
\label{e.H_d_inf}
\end{equation} 
The TFI model is a cornerstone in the theory of QPTs \cite{sachdevbook}: a critical value of the transverse field $g = \Gamma/J = g_c$ separates a low-field ferromagnetic (FM) phase with spontaneously broken symmetry from a high-field quantum paramagnetic (QPM) phase lacking long-range order. Interestingly its infinite-dimensional version, Eq.~\eqref{e.H_d_inf}, has also been often discussed in the theory of spin squeezing \cite{Lawetal2001,Rojo2003,Vidaletal2004,Maetal2009} as well as implemented to dynamically generate spin squeezing in recent atomic physics experiments with spinor gases and trapped ions \cite{Grossetal2010, Bohnetetal2016}. On the contrary the metrological aspects of its finite-$d$ versions have been far less discussed in the literature \cite{Liuetal2013}. Here we focus on the ground-state and finite-temperature properties of the above models making use of its exact solution in $d=1$ and $\infty$, as well as of numerically exact quantum Monte Carlo (QMC) simulations - based on the Stochastic Series Expansion \cite{SyljuasenS2002} (see Supplementary Material - SM - for further details \cite{SM}).   

 \begin{figure}
 \includegraphics[width = 0.9\linewidth]{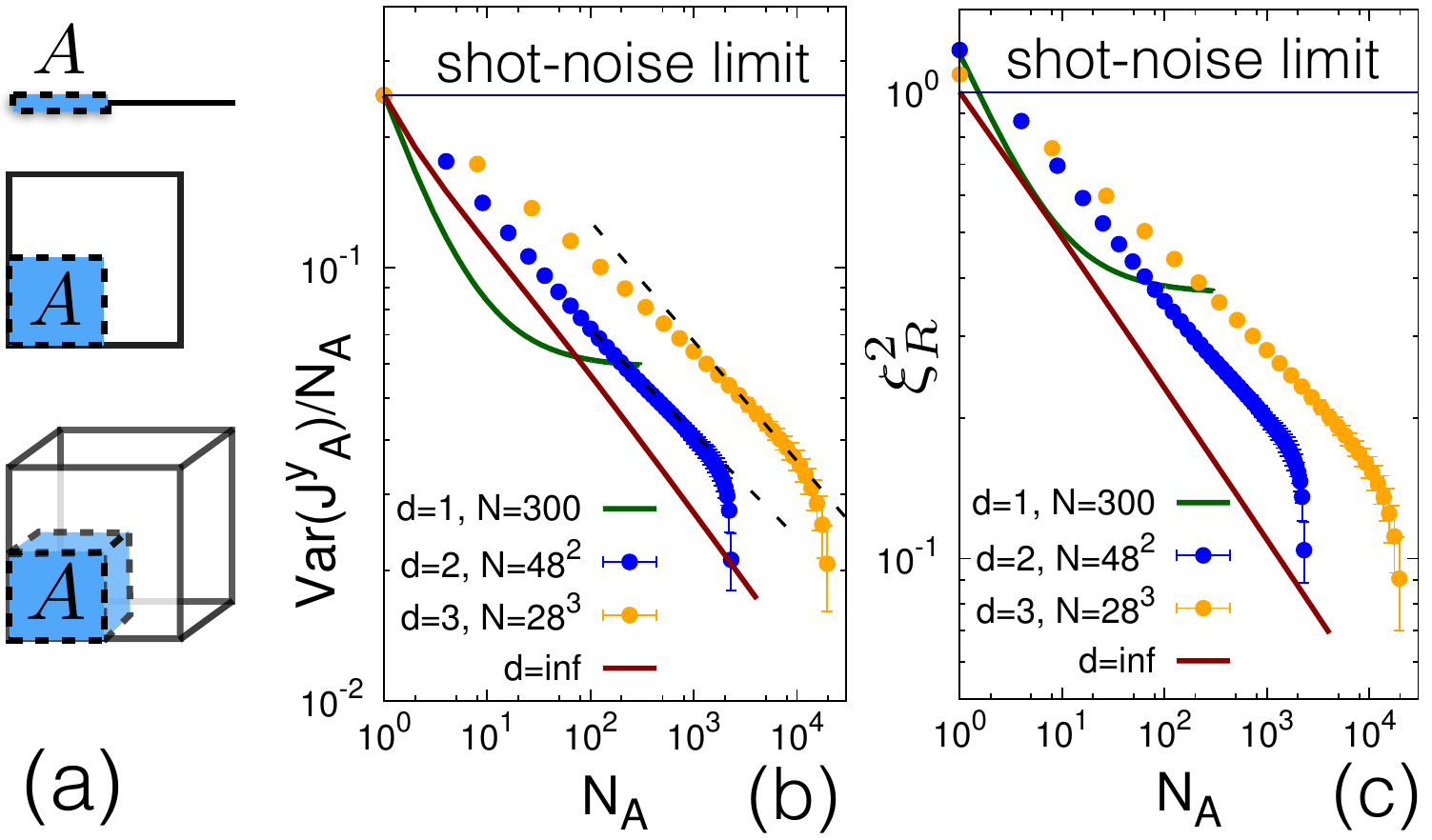}
	\caption{\emph{Squeezing at the quantum-critical point.} (a) Subsystem-$A$ geometries used for the calculation of the scaling of ${\rm Var}(J^y_A)$, namely segments, squares and cubes in $d=1$, 2 and 3, respectively; (b) scaling of the variance of the subsystem collective spin component $J^y_A = \sum_{i\in A} S_i^y$ for subsystems $A$ depicted in (a) in $d=1$, 2 and 3. In the $d=\infty$ case $A$ is not a subsystem but a whole system of size $N_A$. The data are taken at $g=0.6$ and $T=0$ ($d=1$), $g=g_c$ and $T/J = 1/144$ ($d=2$), $g=g_c$ and $T/J = 1/56$ ($d=3$), and $g=g_c$ and $T=0$ ($d=\infty$).  Dashed lines are power-law fits to the form $a\times (N_A)^{\zeta'}$  with $\zeta'\approx 0.24$ for $d=2$ and $0.28$ for $d=3$. The solid line indicates the shot-noise limit ${\rm Var}(J^y_A)/N_A = 1/4$; (c) scaling of the spin-squeezing parameter $\xi_R^2$. Same significance of symbols as in panel (b) (except for $d=1$, where the data refer to $g=0.62$).}
	\label{f.QCsqueezing}
      \end{figure}
      
\emph{Quantum Fisher Information vs. squeezing.} Modeling the interferometer with a unitary transformation $e^{i\phi O}$, the minimal uncertainty on the estimation of the phase $\phi$ is provided by the quantum Fisher Information (QFI) related to the generator $O=\sum_{i=1}^N O_i$ via the quantum Cram\'er-Rao bound \cite{PezzeS2014}
 \begin{equation}
  (\delta \phi)^2 \geq \frac{1}{{k\rm QFI}(O)}  = \frac{\chi^2}{kN} ~;
  \label{e.QCRbound}
  \end{equation}
 the QFI is defined as ${\rm QFI}(O) = \sum_{nm} (p_n-p_m)^2 |\langle m | O | n \rangle|^2/(p_n+p_m)$, where $|n\rangle$ ($|m\rangle$) are eigenstates of the density matrix $\rho$ with eigenvalues $p_n$ ($p_m$); and $k$ is the number of independent measurements performed. A factor $\chi^2 = N/{\rm QFI}(O) <1$ witnesses a metrological gain with respect to the shot-noise limit, as well as the presence of entanglement \cite{Hyllusetal2012, Toth2012} \footnote{We assume that $O_i$ has a spectrum with unit width.}. For a pure state ${\rm QFI}(O) = 4 {\rm Var}(O)$; hence, choosing $O$ as the (macroscopic) order parameter of a QPT, one can exploit its critical super-extensive fluctuations, ${\rm Var}(O) \sim N^{1+\zeta}$ ($\zeta > 0$), to achieve sub-shot-noise precision, namely $(\delta \phi)^2 \sim N^{-1-\zeta}$ and $\chi^{2} \sim N^{-\zeta}$. For the TFI model in dimensions $d\leq d_c= 3$,  $O$ is the $z$-component of the collective spin ${\bm J} = \sum_i {\bm S}_i$, and $\zeta = \frac{1}{d} \left(\frac{2 - \eta}{z}-1 \right) = \frac{1 - \eta}{d}> 0$, where $\eta$ and $z (=1)$ are the correlation function and dynamical critical exponents of the QPT, respectively. Above the upper critical dimension $d>d_c(=3)$, $\zeta$ takes the above form with $d_c$ instead of $d$ \cite{Botetetal1982}, while $\eta = 0$.
 
 This result already embodies the metrological interest of QCPs, but it is otherwise silent about the specific measurement which is able to enjoy the quantum-critical metrological gain witnessed by the QFI. Of much more immediate utility is instead the Heisenberg's uncertainty principle for the collective spin
 \begin{equation}
 {\rm Var}(J^y) \geq \frac{\langle J^x \rangle^2}{4{\rm Var}(J^z)} 
 \end{equation}
 which, at the QCP, allows to conclude that ${\rm Var}(J^y) \geq {\cal O}(N^{1-\zeta})$, namely the lower bound on the variance of the $J^y$ operator acquires a sub-extensive scaling at criticality. Similarly the spin-squeezing parameter \cite{Winelandetal1994}
 \begin{equation}
 \xi_R^{2} = \frac{ N~ {\rm Var}(J^y)}{\langle  J^x \rangle^2 }
 \end{equation}
 (which, when smaller than one, expresses the metrological gain in Ramsey interferometry with respect to uncorrelated states, and also witnesses entanglement \cite{Sorensenetal2001}) acquires a vanishing lower bound at criticality, $\xi_R^{2} \geq N/[4 {\rm Var}(J^z)] \sim {\cal O}(N^{-\zeta})$. This lower bound can also be predicted via the more general inequality $\xi_R^{2} \geq \chi^2$ \cite{PezzeS2009}. The critical scaling of the lower bound on $\xi_R^2$ and ${\rm Var}(J^y)$ is suggestive of the possibility to observe quantum-critical scaling of spin squeezing; but only an explicit microscopic calculation can test whether critical squeezing is indeed achieved or not. 
 
  \begin{figure}[ht!]
 \includegraphics[width = 0.8\linewidth]{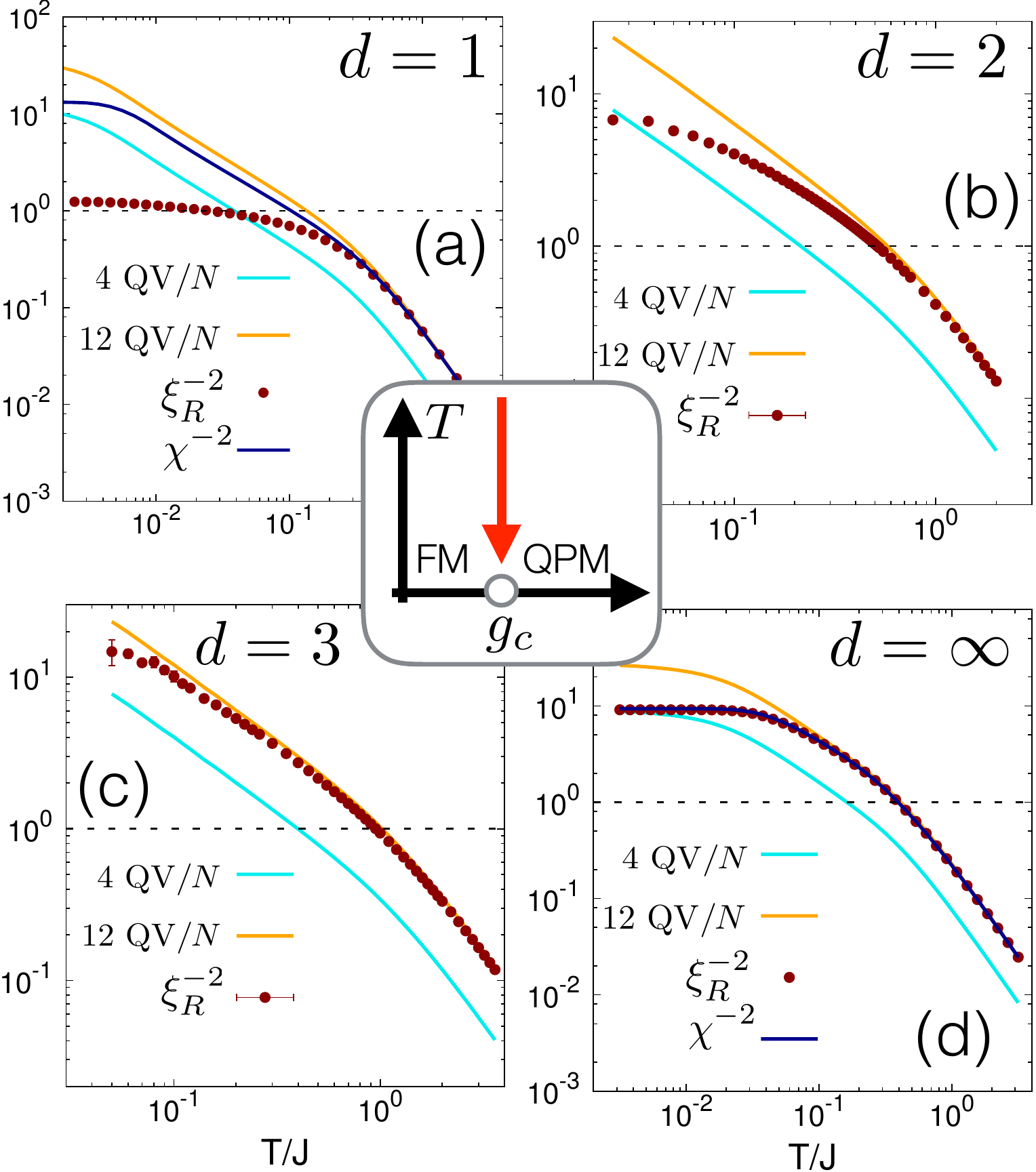}
	\caption{\emph{Quantum correlations along the quantum-critical trajectory.} Squeezing parameter $\xi_R^2$, $\chi^{-2}$ parameter and its bounds provided by the quantum variance for the TFI model as a function of temperature at $g=g_c$: (a) $d=1$, $N=50$; (b) $d=2$, $N = 64^2$; (c) $d=3$, $N=28^3$; (d) $d=\infty$, $N=1000$.}
	\label{f.QCtrajectory}
      \end{figure}

 \emph{Quantum critical squeezing.} Our joint exact/numerical study of the ground-state scaling of $J^y$ fluctuations shows a very complex and intriguing picture upon varying the number of dimensions - see Fig.~\ref{f.QCsqueezing}. 
The case of $d=\infty$ is exactly solved by writing the Hamiltonian in the $|S;M\rangle$ basis of eigenstates of ${\bm J}^2$ and $J^z$, and diagonalizing it in each $S$ sector separately. There at the critical point $g_c=1$ one observes numerically that $\xi_R^{2} \simeq \chi^2 \sim N^{-1/3}$ (as already noticed in previous works \cite{Vidaletal2004,DusuelV2004,Maetal2009,Pezzeetal2016}): namely the ground state in $d=\infty$ is a minimal uncertainty state, realizing the maximum quantum-critical squeezing authorized by Heisenberg's inequality. This can be understood using elementary quantum mechanics, as in the vicinity of the QCP (but strictly speaking not at the QCP) a Holstein-Primakoff transformation maps the model onto a collection of harmonic oscillators, admitting a minimum-uncertainty ground state \cite{Pezzeetal2016, SM}. 

On the opposite side of the spectrum lies the case of $d=1$, whose exact solution, based on Jordan-Wigner mapping onto free fermions \cite{Pfeuty1970}, shows that ${\rm Var}(J^y)$ at the critical point $g_c=1/2$ exhibits a conventional volume-law scaling. Therefore squeezing, albeit present (namely $\xi_R^2 <1$, with a minimum at $g\approx 0.62 > g_c$, and a minimum ${\rm Var}(J^y)$ at $g\approx 0.6$), does not show any sign of quantum critical scaling - as already remarked in Ref.~\cite{Liuetal2013}. This observation is in stark contrast with the $\chi^2$ factor, rapidly scaling to zero as $N^{-3/4}$ ($\eta = 1/4$). Hence conventional Ramsey interferometry is far from being the optimal protocol exploiting the significant metrological potential of the QCP in the 1d TFI model.  

The above results, which were already partly known in the literature \cite{Vidaletal2004,DusuelV2004,Liuetal2013}, are interpolated in a very non-trivial way in the intermediate cases $1 < d < \infty$ (lacking an exact solution). There our QMC results show that squeezing progressively acquires quantum critical scaling, yet generically with a \emph{different} exponent than that predicted by the scaling of the Heisenberg's bound, namely $\xi_R^{-2} \sim N^{-\zeta'} $with $0 < \zeta' < \zeta$. In particular, in $d=2$ at $g_c = 1.52219...$ \cite{BloteD2002} we observe that $\zeta' = 0.24(2)  < \zeta = 0.4818...$ (using the exponents of the 3$d$ Ising universality class \cite{PelissettoV2002}); while at the mean-field transition in $d=3$ ($g_c = 2.579....$  \cite{BloteD2002}) we observe $\zeta' = 0.28(2)  < \zeta = 1/3$. In the case $d=3$ we find that our results are still strongly affected by finite-size effects, and we cannot exclude that calculations on larger system sizes may give $\zeta = \zeta' $, while this appears to be very unlikely in the case $d=2$. Hence, for $d=1$ and 2, quantum-critical scaling of spin squeezing introduces a critical exponent $\zeta'$ which, to our knowledge, is a yet unknown combination of the critical exponents of the QCP. 
Rather counterintuitively these results also establish that, the number $N$ of spins being held fixed, the \emph{a priori} metrological potential offered by the QCP (set by the QFI via Eq.~\eqref{e.QCRbound}) decreases with the number of dimensions (as $\zeta$ decreases); but the quantum-critical scaling of squeezing becomes more pronounced (as $\zeta'$ increases with $d$).   

\emph{Quantum correlations and squeezing along the QC trajectory.}  We now turn to the finite-temperature case, which is the most relevant situation from the point of view of  potential experimental implementations. A realistic experimental situation involves the system being prepared with $g \gg 1$ - namely in a coherent spin state $\otimes_{i=1}^N | \uparrow_x \rangle_i$ - and then adiabatically transformed by lowering the  $g$ ratio towards the critical $g_c$ value. Inevitable deviations from adiabaticity will produce an equilibrium state at finite temperature at the end of the $g$ ramp. We then ask the question: how much of the remarkable metrological properties of the ground state survive at finite temperatures in the vicinity of the QCP? 
  
  We start addressing this question by exploring the evolution of metrologically relevant observables along the so-called quantum-critical trajectory (sketch in Fig.~\ref{f.QCtrajectory}), namely by scanning the temperature at $g=g_c$. 
Figs.~\ref{f.QCtrajectory}(a-d) show the temperature dependence of the squeezing parameter, along with that of the $\chi^{-2}$ parameter (when calculable, namely for $d=1$ \cite{Haukeetal2016} and $\infty$), as well as the quantum variance (QV) of the order parameter, introduced by us in Ref.~\cite{FrerotR2016}. The latter is defined as
\begin{equation}
{\rm QV}(J^z) = \langle (J^z)^2 \rangle - k_B T \int_0^{(k_BT)^{-1}} d\tau ~ \langle J^z(\tau) J^z(0) \rangle  
\end{equation}
where $J^z(\tau) = e^{\tau \cal H} J^z e^{-\tau \cal H}$. The QV is known \cite{FrerotR2016,FrerotRunp} to tightly bound the QFI, and hence the $\chi^2$ parameter, as
\begin{equation}
 \frac{{\rm QV}(J^z)}{N} \leq \frac{1}{4} ~\chi^{-2}\leq  3~ \frac{{\rm QV}(J^z)}{N}~.
\end{equation}  
As a consequence $\xi_R^{-2} \leq  \chi^{-2} \leq 12~ {\rm QV}(J^z)/N$.  These bounds to the $\chi^{-2}$ parameter turn out to be extremely useful, because: 1) they are thermodynamical quantities, generically computable with large-scale numerics such as the QMC adopted here, while the QFI (contained in $\chi$) is not, unless one has access to the exact solution of the model \cite{Haukeetal2016}; 2) the joint upper bound to the squeezing parameter and the $\chi^{-2}$ one offered by the QV allows to probe directly how close $\xi_R$ and $\chi$ are, even if one does not know $\chi$ -- indeed if $\xi_R^{-2}$ approaches $12~ {\rm QV}/N$ we know for sure that $\chi^{-2}$ is tightly sandwiched in between. 
We observe that in all dimensions $\xi_R^{-2}$ saturates its upper bound (and therefore coincide with $\chi^{-2}$) for sufficiently high temperatures, namely the QFI and the squeezing parameter contain the same information. But the lower the dimension, the higher the temperature at which the two quantities start to deviate - and particularly so in $d=1$, as $\chi^{-2}$ displays a power-law divergence as $T\to 0$ (consistent with QC behavior \cite{Haukeetal2016, FrerotRunp}), while $\xi_R^{-2}$ does not diverge. For $d>1$ $\xi_R^{-2}$ is seen to exhibit QC temperature scaling, consistent with its divergence at $T=0$, but with a seemingly different power law with respect to the one of $\chi^{-2}$ and of the QV (which exhibits the same divergence as the QFI \cite{FrerotRunp}); yet already in $d=3$ the squeezing parameter and the $\chi^2$ parameter remain extremely close to each other down to very low temperatures $T \sim 10^{-1} J$.  Finally for $d=\infty$,  $\xi_R^{2}$  and  $\chi^2$ are seen to coincide at any temperature, and this despite the strong finite-size effects that infinite-range interactions entail.  
   \begin{figure}
 \includegraphics[width = 0.9\linewidth]{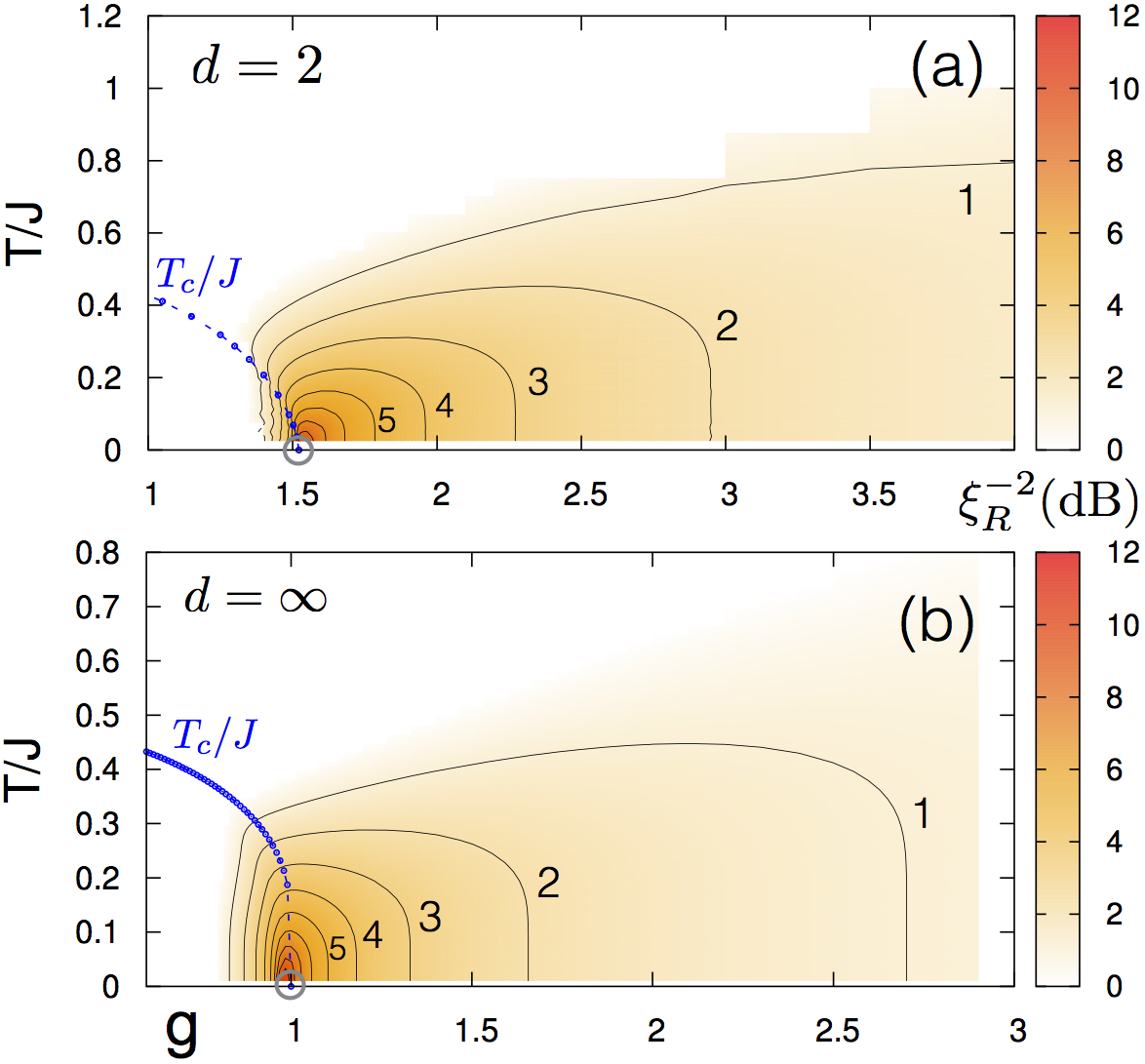}
	\caption{\emph{Squeezing around the QCP.} Squeezing parameter $\xi_R^{-2}$ in dB (across the phase diagram of the TFI model close to the QCP: (a) $d=2$, $N=64^2$; (b) $d=\infty$, $N=500$. The gray circle marks the QCP, and the dashed blue lines indicate the critical temperatures $T_c$ on the ordered side ($T_c$ values for $d=2$ from Ref.~\cite{HesselmannW2016}, and for $d=\infty$ from Ref.~\cite{dasetal2006}). In the white region $\xi_R^{-2}  \leq 1$ (absence of squeezing).}
	\label{f.Tsqueezing}
      \end{figure} 
  
 \emph{Finite-temperature squeezing around the critical point.} Finally, to demonstrate the potential metrological utility of the equilibrium physics close to the QCP, we map out the squeezing parameter in the temperature-field plane. Fig.~\ref{f.Tsqueezing} shows $\xi_R^{-2}$ as a function of the field and temperature in the case of $d=2$ and $\infty$ (analogous figures for $d=1$ and 3 are shown in the SM \cite{SM}). It is very remarkable to observe that the very existence of squeezing, $\xi_R<1$, is essentially induced in the model by the existence of the QCP. Indeed for $g\to \infty$ the ground state is a coherent spin state with $\xi_R = 1$, and squeezing is not produced at finite temperature either. The introduction of spin-spin interactions $g < \infty$ produces correlations, entanglement as well as squeezing in the ground state -- a perturbative calculation \cite{SM} shows that $\xi_R^2 = 1-d/(2g) + {\cal O}(g^{-2})$ -- and $\xi_R$ decreases monotonically upon decreasing $g$ towards the QCP. Such ground-state squeezing is protected at finite temperature by the existence of the spectral gap, controlled by the field (and linear in $\Gamma$ at large $\Gamma$). Upon approaching the QCP the gap closes, but ground-state squeezing becomes critical (in $d>1$) and as a consequence it remains sizable at finite temperature (up to $T/J \sim 0.5$). Once the QCP is crossed, squeezing is quickly lost as one enters the ordered phase - the finite-size ground state for $g \ll g_c$ is a Schr\"odinger's cat state with no squeezing.  
 
\emph{Conclusions and perspectives.} In this work we have unveiled the interest of using equilibrium quantum many-body states lying in the vicinity of a quantum-critical point (QCP) as input states for interferometric measurements which beat the shot-noise limit. We have revealed that extreme spin squeezing - diverging with system size - appears at the QCP of the quantum Ising model in $d>1$, and that very strong squeezing - associated with equally strong quantum correlations - survives up to sizable temperatures above the QCP.  In particular the precision of standard Raman interferometry interrogating the collective spin of the output state nearly saturates the quantum Cram\'er-Rao bound down to low temperatures in $d=3$ and higher, showing that the quantum correlations of QCPs can be potentially exploited in current metrological setups such as atomic clocks. The metrological potential of QCPs in lower dimensions can instead only be exploited via more complex observables than the collective spin, signaling the non-Gaussian nature of the corresponding states \cite{Strobeletal2014} - work is in progress to identify such observables. Our findings are immediately relevant to quantum simulation setups realizing the quantum Ising model and its quantum phase transition -- namely trapped ions \cite{Bohnetetal2016}, Rydberg atoms \cite{Labuhnetal2016}, or ultracold binary atomic mixtures \cite{Sabbatinietal2011, FrerotRunp} -- suggesting that quantum simulators of quantum critical phenomena can potentially find an application as quantum sensors.  

\textit{Acknowledgments.} We thank A. Ran\c con for useful discussions, S. Hesselmann for sharing the data of Ref.~\cite{HesselmannW2016}. This work is supported by ANR (``ArtiQ" project).

\newpage

\begin{center}
{\bf Supplementary Material } \\
{\bf \emph{``Quantum critical metrology"}}
\end{center}

\begin{center}
{\bf Details of the exact and numerically exact calculations}
\end{center}

The $d=1$ transverse-field Ising (TFI) model with \emph{open} boundary conditions is solved exactly via a Jordan-Wigner transformation, mapping it onto a chain of free fermions \cite{Pfeuty1970}. The fermionic density provides the transverse magnetization $\langle J^x \rangle $, while the correlation function $\langle S_i^y S_j^y \rangle$ can be expressed as the Pfaffian of an antisymmetric matrix \cite{derzhkoK1997}, and then used to calculate ${\rm Var}(J^y)$. The calculation of the ${\rm QFI}(J^z)$ is more intricate, but made possible thanks to the link with the dynamical structure factor established by Ref.~\cite{Haukeetal2016}. To this goal we calculate the time-dependent correlation function $\langle S_i^z(t) S_j^z(0) \rangle$ - which can be expressed as a  Pfaffian as well \cite{derzhkoK1997} - and then take the Fourier transform to obtain the dynamical structure factor. Efficient calculation of Pfaffians is achieved in Python via the library published in Ref.~\cite{wimmer2012}. The quantum variance is obtained by calculating the imaginary-time correlation function $\langle S_i^z(\tau) S_j^z(0) \rangle$ averaged over $\tau$ between $\tau= 0$ and $\beta$ (see main text and Ref. \cite{FrerotR2016}). 

The $d=2$ and $d=3$ TFI model with \emph{periodic} boundary conditions is solved via quantum Monte Carlo simulations based on the Stochastic Series Expansion representation. The SSE formulation we use is slightly unusual in that, unlike the algorithm in Ref.~\cite{Sandvik2003}, the quantization axis is chosen along the field axis ($x$). This choice produces directed-loop updates which, unlike in the above-cited algorithm, are not confined to single sites. This aspect allows us to reconstruct the  $\langle S_i^y S_j^y  \rangle $ correlation function during the directed-loop update \cite{SyljuasenS2002}, producing the rich statistics necessary to probe the weak fluctuations of $J^y$ - the central focus of this work. 

The $d=\infty$ TFI model is solved via exact diagonalization in the collective-spin basis $|S;M\rangle$, which was carried out in all $S$ sectors. Calculations of the collective-spin averages and fluctuations are straightforward, including that of the QFI, based on its very definition (see main text). The quantum variance is calculated thanks to a similar formula given in Ref. \cite{FrerotR2016}.

\begin{center}
{\bf Holstein-Primakoff treatment of the $d=\infty$ transverse-field Ising model}
\end{center}

 We briefly review the Holstein-Primakoff (HP) approach to the infinite-range Ising model, with the aim of exposing the connection of the latter with the quantum harmonic oscillator.  
In the $S\to\infty$ ($N\to\infty$) limit, the ground state of the infinite-range Ising model can be solved by treating it as a classical spin, with orientation ${\bm J}= S (\cos\theta, 0,\sin\theta)$, where $\theta = \arcsin g$ for $g \leq 1$ and $\theta = \pi/2$ for $g>1$. Defining the quantization axis along the classical spin orientation, and using the (linearized) HP transformation \cite{Mattis2006}, the collective spin is mapped onto bosonic operators $a$, $a^{\dagger}$ ($[a,a^{\dagger}]=1$) dimensionless canonical position/momentum variables with commutation relation $[x,p] = i$ as
\begin{eqnarray}
J^x & = & \sin\theta \left ( S- a^{\dagger} a \right) + \cos\theta \sqrt{\frac{S}{2}}~ (a+a^{\dagger}) + {\cal O}\left (\frac{1}{\sqrt{S}}\right) \nonumber \\
J^y & = & \sqrt{\frac{S}{2}}\frac{a-a^{\dagger}}{i}   + {\cal O}\left (\frac{1}{\sqrt{S}}\right) \\
J^z & = &  \cos\theta \left ( S- a^{\dagger} a \right) - \sin\theta \sqrt{\frac{S}{2}}~ (a+a^{\dagger}) + {\cal O}\left (\frac{1}{\sqrt{S}}\right) \nonumber
\end{eqnarray}  
and the Hamiltonian takes a quadratic form 
\begin{equation}
{\cal H} = E_0(g) + J f(g) \left[ \frac{p^2}{2} + \frac{1}{2} \omega^2(g) ~x^2 + {\cal O}(S^{-1}) \right] 
\end{equation}
where $f(g) = 1$ for $g\leq1$ and $f(g) = g$ for $g>1$; $\omega^2(g) = 1-g^2$ for $g\leq 1$ and $1-1/g$ for $g>1$; and $E_0(g) = -JS(\cos^2\theta +g \sin\theta) $ is the classical energy - here we are using the fact that, in the ground state, $S=N/2$. The ground state of the system within the linearized HP approach is therefore that of a quantum harmonic oscillator (except at the critical point $g=1$), and in its ground state we straightforwardly obtain that:
\begin{eqnarray}
\langle J^x \rangle  &= &  \sin\theta~ S \left[ 1 + {\cal O}\left (\frac{1}{\sqrt{S}}\right) \right] \nonumber \\
{\rm Var}(J^y) & = & \frac{S}{2} \left[ 1 + {\cal O}(S^{-1})  \right]  \nonumber \\
{\rm Var}(J^z) & = & \sin^2 \theta ~\frac{S}{2} \left[1 + {\cal O}\left (\frac{1}{\sqrt{S}}\right) \right]~.
\end{eqnarray}
As a consequence we observe that
\begin{equation}
{\rm Var}(J^y) {\rm Var}(J^z) = \frac{ \langle J^x \rangle^2}{4} ~\left(1 + {\cal O}\left (\frac{1}{\sqrt{S}}\right) \right)
\end{equation}
namely, in the thermodynamic limit $S=N/2 \to\infty$ the ground state is a minimum uncertainty state of the collective spin.

  \begin{figure*}[ht!]
 \includegraphics[width = 0.9\linewidth]{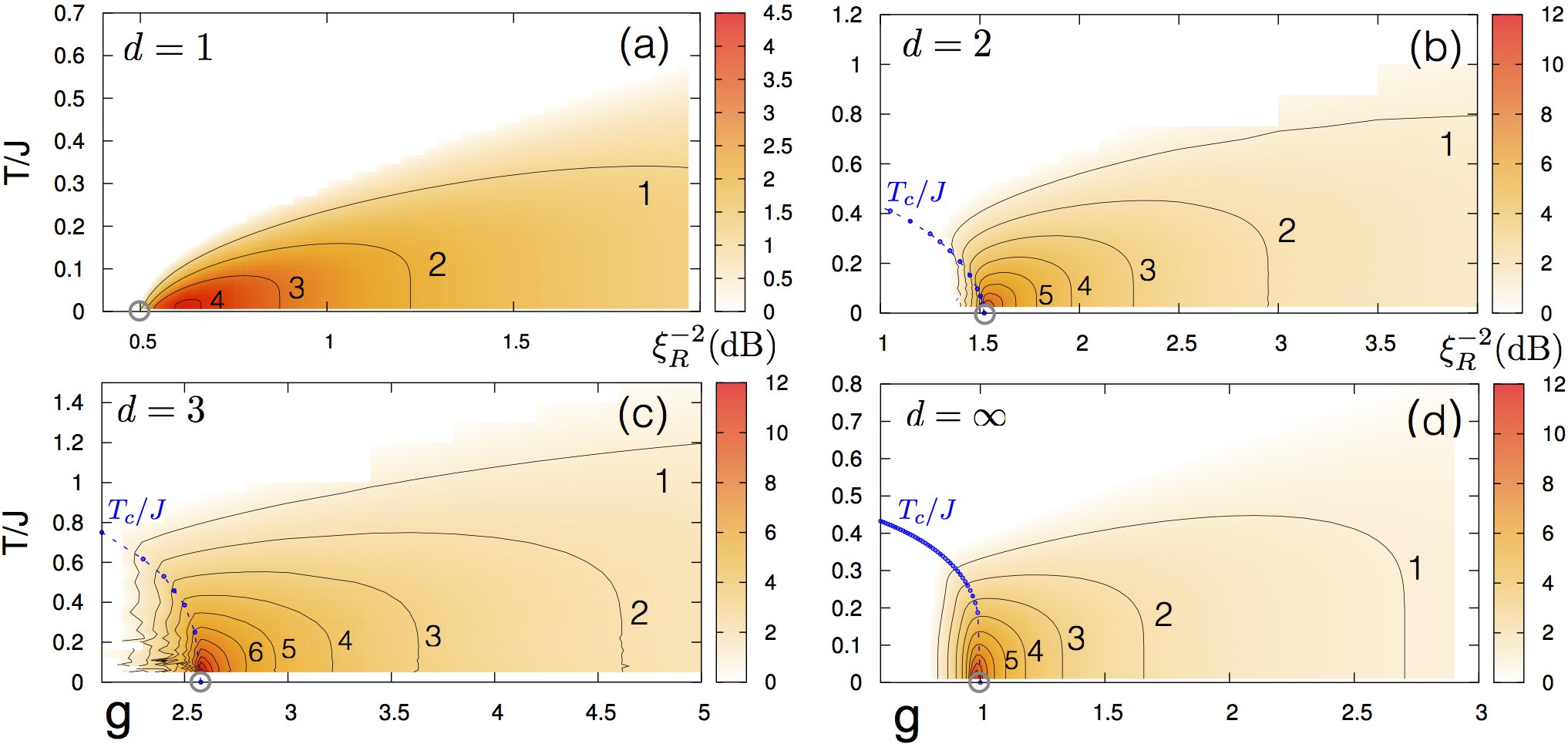}
	\caption{Squeezing parameter $\xi_R^{-2}$ across the phase diagram of the TFI model close to the QCP: (a) $d=1$, $N=50$; (b) $d=2$, $N=64^2$; (c) $d=3$, $N=28^3$; (b) $d=\infty$, $N=500$. The gray circle marks the QCP, and the dashed blue lines indicate the critical temperatures on the ordered side. In the white region $\xi_R^{-2}  \leq 1$ (absence of squeezing).}
	\label{f.Tsqueezing_ALL}
      \end{figure*}

  \begin{figure}[ht!]
 \includegraphics[width = 0.9\linewidth]{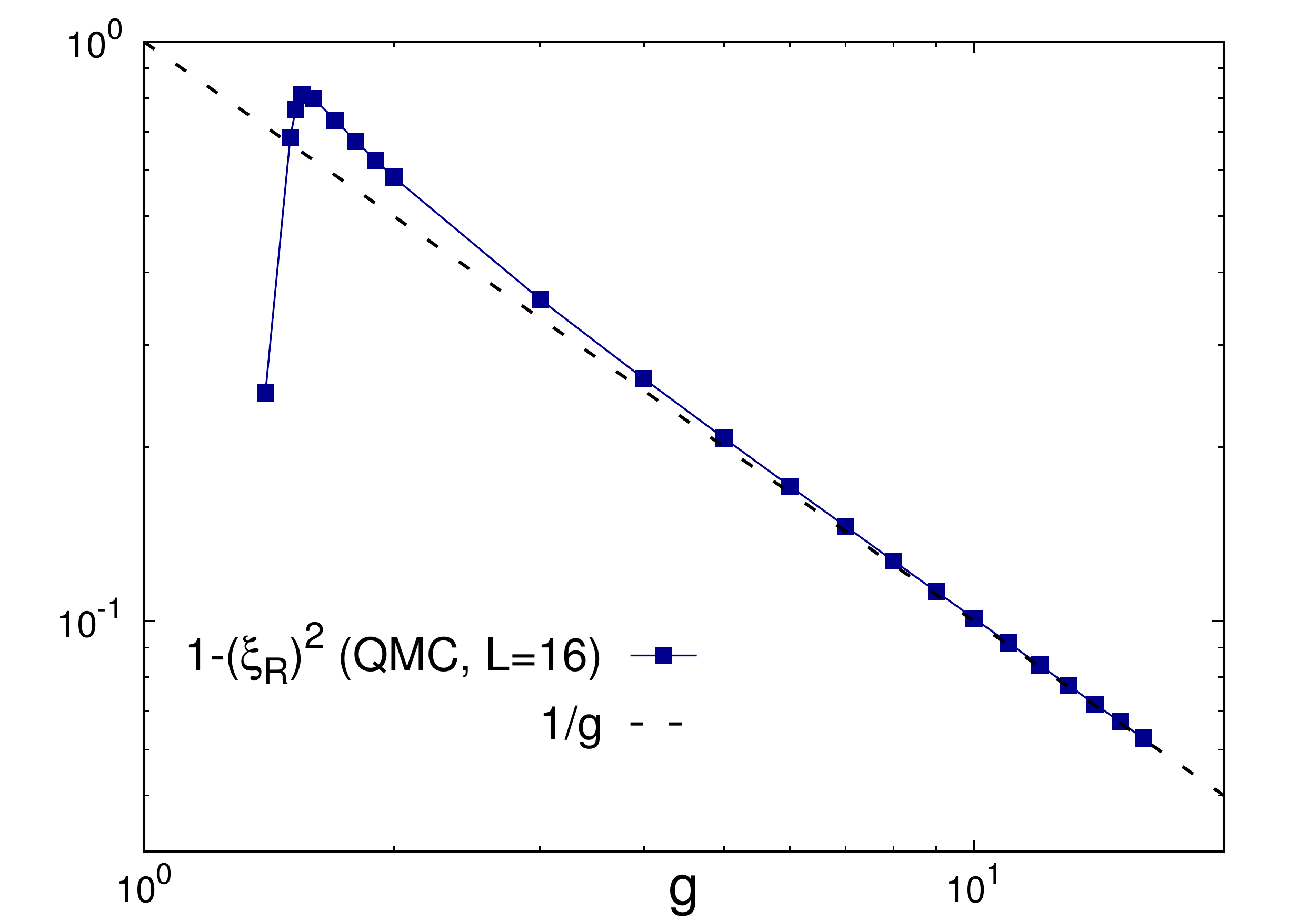}
	\caption{Squeezing parameter $1-\xi_R^2$ as function of $g$ in the large-$g$ limit. The dots correspond to a QMC simulation on a $L\times L$ square lattice with $L=16$ at temperature $T = 1/L$; the dashed line is the first-order perturbative prediction (see text).}
	\label{f.largeg}
      \end{figure}

\begin{center}
{\bf Finite-temperature squeezing from $d=1$ to $d=\infty$}
\end{center}

Fig.~\ref{f.Tsqueezing_ALL} shows the inverse squeezing parameter ($\xi_R^{-2}$) in dB across the temperature-field phase diagram of the TFI model in the vicinity of the quantum critical point. We show our results for dimensions $d=1$, 2, 3 and $\infty$ (for the sake of completeness we reproduce again the diagrams for $d=2$ and $\infty$ which are already to be found in the main text). All the data are for finite-size systems, but the systems sizes for $d\leq 3$ are sufficiently large for finite-size effects on $\xi_R^{-2}$ to be essentially negligible. 

We observe that for $d\geq 2$ the quantum-critical point realizes the maximum squeezing, with $\xi_R^{-2}$ diverging at $T=0$, and remaining sizeably higher than $1$  up to high temperatures  $T \sim J$. In particular squeezing is present over a large portion of the phase diagram on the disordered side ($g>g_c$) , whereas on the ordered side ($g<g_c$) squeezing disappears rapidly as one enters the ordered phase for $T<T_c$. 
On the other hand, in $d=1$ the QCP displays very little squeezing, whereas the maximum squeezing is realized at a field strength $g \approx 0.6 > g_c =1/2$.

\begin{center}
{\bf Squeezing at strong fields ($g\to \infty$): perturbative calculation}
\end{center}

 We develop an elementary perturbation theory calculation in the strong-field limit $g = \Gamma/J \to \infty$ for the TFI model in $d \leq \infty$. Writing the Hamiltonian as 
 \begin{equation}
 \frac{{\cal H}}{\Gamma} = \frac{\cal K_{\rm I}}{g} -  J^x 
\end{equation}
with ${\cal K}_{\rm I} = \sum_{\langle ij \rangle} S_i^z S_j^z$ representing the Ising coupling and $J^x = \sum_i S_i^x$ representing the transverse magnetization. We treat the $-J^x$ term as the unperturbed Hamiltonian, admitting the unperturbed ground state $|\psi_0\rangle = \otimes_l |\uparrow_x \rangle_l$ (the coherent spin state); and the  ${\cal K}_{\rm I}/g$ term as a perturbation. The first-order perturbed ground state takes the form 
\begin{eqnarray}
|\psi\rangle & = & {\cal N}^{-1} \Big (|\psi_0\rangle  \\
&- & \frac{1}{8g} \sum_{\langle ij\rangle} |\uparrow_x \rangle_1\otimes ... |\downarrow_x \rangle_i... |\downarrow_x \rangle_j ... |\uparrow_x \rangle_N + {\cal O}(g^{-2}) \Big) \nonumber
\end{eqnarray}
where ${\cal N} = \sqrt{1+ \frac{Nz}{16 g} + {\cal O}(g^{-2})}$ is the normalization factor, and $z$ is the coordination number of the lattice ($z=2d$ for hypercubic lattices). The resulting correlation function for the $S^y$ spin components takes then the form
\begin{equation}
\langle \psi | S_i^y S_j^y |\psi \rangle = \frac{1}{4} \delta_{ij} - \frac{1}{16 g} \delta_{\langle ij \rangle} + {\cal O}(g^{-2})
\end{equation}
where $\delta_{\langle ij \rangle}=1$ if $i$ and $j$ are nearest neighbors, and zero otherwise. 
As a consequence, by integration of the correlation function we readily obtain 
\begin{equation}
{\rm Var}(J^y) = \frac{N}{4} \left (1 - \frac{z}{4g} \right) + {\cal O}(g^{-2})
\end{equation}
Similarly one obtains that $\langle J^x \rangle = N/2 + {\cal O}(g^{-2})$. As a result
\begin{equation}
\xi_R^2 = \frac{N ~{\rm Var}(J^y)}{\langle J^x \rangle^2} = 1 - \frac{z}{4g} +  {\cal O}(g^{-2})
\end{equation}
as stated in the main text. Fig.~\ref{f.largeg} shows that this behavior is indeed verified by our QMC results for the case $d=2$. 

In the case $d=\infty$, $z =N$ and $g\to Ng$, so that 
 \begin{equation}
\xi_R^2 (d=\infty) = 1 - \frac{1}{4g} +  {\cal O}(g^{-2})~.
\end{equation}

\bibliography{QCmetro_1.2.bib}

\end{document}